\documentclass[twocolumn,showpacs,preprintnumbers,amsmath,amssymb]{revtex4-1}

\usepackage{afterpage}
\usepackage[dvipdfmx]{graphicx}
\usepackage{color,bm}

\begin{document}

\title{Triangular solution to the planar elliptic three-body problem 
in the parametrized post-Newtonian formalism}
\author{Yuya Nakamura}
\email{nakamura@tap.st.hirosaki-u.ac.jp}
\author{Hideki Asada} 
\email{asada@hirosaki-u.ac.jp}
\affiliation{
Graduate School of Science and Technology, 
Hirosaki University,
Hirosaki 036-8561, Japan} 

\date{\today}

\begin{abstract}
A triangular solution [Phys. Rev. D 107, 044005 (2023)] 
has recently been found 
to the planar circular three-body problem 
in the parametrized post-Newtonian (PPN) formalism, 
for which they focus on a class of fully conservative theories 
characterized by the Eddington-Robertson parameters $\beta$ and $\gamma$. 
The present paper extends the PPN triangular solution 
to quasi-elliptic motion, 
for which the shape of the triangular configuration changes with time 
at the PPN order. 
The periastron shift due to the PPN effects is also obtained. 
\end{abstract}


\maketitle

\section{Introduction}
The three-body problem is among the classical ones in physics, 
which led to a study of the chaos 
\cite{Goldstein}. 
Particular solutions, 
notably
Euler's collinear solution and Lagrange's equilateral one 
\cite{Danby,Marchal} 
represent regular orbits, which have attracted a lot of interest 
e.g. \cite{Asada,Torigoe,Seto,Schnittman,Connors}. 

Nordtvedt \cite{Nordtvedt} pointed out that 
the position of the triangular points 
is very sensitive to the ratio between the gravitational mass 
and the inertial one in gravitational experimental tests, 
though the post-Newtonian (PN) terms are partly considered. 

For the restricted three-body problem in the PN approximation, 
Krefetz \cite{Krefetz} and Maindl \cite{Maindl} 
found the PN triangular configuration for 
a general mass ratio between two masses. 
These studies were extended to 
the PN three-body problem for 
general masses 
\cite{Yamada2010,Yamada2011,Ichita2011,Yamada2012,Yamada2015,Yamada2016}, 
where  
the PN counterparts for 
Euler's collinear \cite{Yamada2010,Yamada2011} 
and Lagrange's equilateral solutions \cite{Ichita2011,Yamada2012}
were found. 
It should be noted that 
the PN triangular solutions are not necessarily equilateral 
for general mass ratios 
and they are equilateral only for either the equal mass case 
or two test masses. 
The stability of the PN solution  and 
the radiation reaction at 2.5PN order 
were also studied \cite{Yamada2015,Yamada2016}. 

In a scalar-tensor theory of gravity, 
a collinear configuration for three-body problem 
was discussed 
\cite{Zhou}. 
In addition to such fully classical treatments, 
a possible quantum gravity correction to the Lagrange points was argued  \cite{Battista2015a,Battista2015b}. 

Moreover, the recent discovery of a relativistic hierarchical triple system 
including a neutron star \cite{Ransom} 
has sparked   
renewed interest in the relativistic three-body problem 
and the related gravitational experiments  
\cite{Archibald,Will2018,Voisin}. 

In the PPN formalism, 
collinear and triangular solutions 
to the planar circular three-body problem 
have recently been found 
\cite{Nakamura2023}, 
where they focus on a class of fully conservative theories characterized by 
the Eddington-Robertson parameters $\beta$ and $\gamma$, 
because the two parameters are the most important ones; 
$\beta$ measures how much nonlinearity there is in the superposition law for gravity 
and $\gamma$ measures how much space curvature is produced by unit rest mass 
\cite{Will,Poisson} . 
See e.g. \cite{Klioner} for the celestial mechanics in this class of PPN theories. 

In the Newtonian gravity, 
triangular solutions are not only 
to the circular three-body problem 
but also to the elliptic one  
\cite{Danby, Roy}. 
Can a (quasi-)elliptic orbit of triangular solutions 
be found in PPN case? 
A point is that the PPN force 
seems to be too complicated 
to admit elliptic orbits for a triple system. 
The main purpose of the present paper is 
to find it in the class of fully conservative theories. 

This paper is organized as follows. 
In Section II, 
basic methods and equations are presented. 
Section III discusses 
the PPN triangular solution to the planar elliptic three-body problem. 
Section V summarizes this paper. 
Throughout this paper, $G=c=1$. 
$A, B$ and $C \in \{1,2,3\}$ label three masses.

\section{Basic methods and equations}
\subsection{Newtonian planar elliptic triangular solution}
Let us begin by briefly summarizing the triangular solution 
to the Newtonian planar elliptic three-body problem 
\cite{Danby,Roy}.
A homothetic solution is possible and it represents the Lagrange equilateral solution 
in elliptic motion. 
See e.g. Section 5 of Reference \cite{Danby} for more detail. 
We shall see that the PN triangular solutions are not necessarily equilateral,
 mainly because of the velocity-dependent force at the PN order 
as shown in Section III.

The equation of motion (EOM) for three masses 
($M_A$ at the position $\bm{R}_A$) reads
\begin{align}
M_A \bm{a}_A 
= -\sum_{B=1}^{N} \frac{M_A M_B}{(R_{AB})^2} \bm{n}_{AB} ,
\label{EOM-Newton}
\end{align}
where 
$ \bm{a}_A$ denotes the acceleration of the $A$-th mass, 
$\bm{R}_{AB} \equiv \bm{R}_A -  \bm{R}_B$, 
$R_{AB} \equiv |\bm{R}_{AB}|$,  
and 
$\bm{n}_{AB} \equiv \bm{R}_{AB}/R_{AB}$. 

By taking the cross product of $\bm{R}_1$ and 
Eq. (\ref{EOM-Newton}) for $A=1$, 
we obtain 
\begin{align}
\bm{R}_1 \times \bm{R}_2 
\left(
\frac{1}{(R_{12})^3} - \frac{1}{(R_{31})^3}
\right)
= 
0 ,
\label{cross-Newton}
\end{align}
where the coordinate center is chosen as the center of mass (COM) 
of $\sum_A M_A \bm{R}_A = 0$.
For a triangular configuration, 
$\bm{R}_1 \nparallel \bm{R}_2$. 
From Eq. (\ref{cross-Newton}), 
we thus obtain $R_{12} = R_{23}$. 
By cyclic arguments, we obtain an equilateral solution 
\cite{Danby,Roy}. 

In elliptic motion, 
the arm length $R_A$ becomes 
$R_1 = a f_N  \sqrt{\nu_2^2 + \nu_2 \nu_3 + \nu_3^2}$, 
$R_2 = a f_N \sqrt{\nu_3^2 + \nu_3 \nu_1 + \nu_1^2}$, 
and 
$R_3 = a f_N \sqrt{\nu_1^2 + \nu_1 \nu_2 + \nu_2^2}$, 
where  
the total mass is $M \equiv \sum_A M_A$,  
the mass ratio is defined as $\nu_A \equiv M_A/M$, 
$a$ is some constant, 
and $f_N$ denotes the dilation factor 
\cite{Danby,Roy,Ichita2011,Yamada2012}. 
In circular motion, $f_N = 1$, 
while  
$f_N$ is a function of time 
in elliptic motion 
\cite{Danby,Roy}.

From the total energy and angular momentum, 
an elliptic orbit is obtained as 
\cite{Danby,Roy}
\begin{align}
f_{\rm{N}}
 = 
 \frac{\mathcal{A}_{\rm{N}} (1 - e_{\rm{N}}^2)}
{1 + e_{\rm{N}} \cos\theta} ,
\label{orbit-Newton}
\end{align}
where 
$\theta$ denotes the true anomaly, 
$e_{\rm{N}}$ is the eccentricity of the elliptic orbit as 
$e_{\rm{N}} = \sqrt{1 + 2 L_{\rm{N}}^2 \mathcal{E}_{\rm{N}}M^{-2} \mu^{-3}}$ 
for the total energy $\mathcal{E}_{\rm{N}}$, 
the total angular momentum $L_{\rm{N}}$ 
and 
$\mu \equiv M (\nu_1\nu_2 + \nu_2\nu_3 + \nu_3\nu_1)$, 
and 
$\mathcal{A}_{\rm{N}} \equiv - \mu M/(2 a \mathcal{E}_{\rm{N}})$.
Here, 
$\theta = 0$ is chosen as the periastron. 

For the simplicity,  
we refer to $A \equiv a \mathcal{A}_{\rm{N}}$ as the semi-major axis and 
$P \equiv a \mathcal{A}_{\rm{N}} (1 - e_{\rm{N}}^2) $ 
as the semi-latus rectum. 
For instance, the semi-major axis for the elliptic orbit of $M_1$ is 
$a \mathcal{A}_{\rm{N}} \sqrt{\nu_2^2 + \nu_2 \nu_3 + \nu_3^2}$.

From the total angular momentum, 
the angular velocity $\omega_{\rm{N}}$ of the triangular configuration 
is obtained as 
\begin{align}
\omega_{\rm{N}} 
=& ( 1 + e_{\rm{N}} \cos \theta )^2\sqrt{\frac{M}{P^3}}  . 
\label{omega-Newton}
\end{align}

All the above relations are reduced to Keplerian orbits 
in the restricted three-body problem (e.g. $\nu_3 \to 0$).


\subsection{EOM in the PPN formalism}
In a class of fully conservative theories 
including only the Eddington-Robertson parameters $\beta$ and $\gamma$,  
the PPN EOM becomes 
\cite{Will,Poisson} 
\begin{align}
\bm{a}_A =& - \sum_{B\neq A} \frac{M_B}{R_{AB}^2} \bm{n}_{AB} 
          \notag\\
          & - \sum_{B\neq A } \frac{M_B}{R_{AB}^2} \bigg\{ \gamma v_A^2 - 2(\gamma + 1)(\bm{v}_A \cdot \bm{v}_B) \nonumber\\
        &~~~~~ + (\gamma + 1)v_B^2 
         - \frac{3}{2} (\bm{n}_{AB} \cdot \bm{v}_B)^2      
         - \bigg(2\gamma + 2\beta +1  \bigg) \frac{M_A}{R_{AB}} 
         \notag\\
         &~~~~~
       - 2(\gamma +\beta) \frac{M_B}{R_{AB}} 
         \bigg\} \bm{n}_{AB} \nonumber\\  
       & + \sum_{B\neq A} \frac{M_B}{R_{AB}^2} \bigg\{\bm{n}_{AB} \cdot [2(\gamma +1) \bm{v}_A - (2\gamma+1) \bm{v}_B]
         \bigg\}(\bm{v}_A - \bm{v}_B) \nonumber\\       
      & + \sum_{B \neq A} \sum_{C\neq A, B} \frac{M_B M_C}{R_{AB}^2} 
      \bigg[ \frac{2(\gamma +\beta)}{R_{AC}} +  \frac{2\beta -1}{R_{BC}} 
      \notag\\
      &~~~~~~~~~~~~~~~~~~~~~~~~~~~~~~ 
      -\frac{1}{2}  \frac{R_{AB}}{R_{BC}^2} (\bm{n}_{AB}\cdot\bm{n}_{BC}) 
      \bigg] \bm{n}_{AB} \nonumber\\
    &- \frac{1}{2} (4\gamma +3 ) 
     \sum_{B\neq A} \sum_{C\neq A,B} \frac{M_B M_C}{R_{AB} R_{BC}^2} \bm{n}_{BC}
          + O(c^{-4}) , 
\label{EOM-PPN}
\end{align}
where 
$ \bm{v}_A$ denotes the velocity of the $A$-th mass.

\begin{figure}
\includegraphics[width=7.5cm]{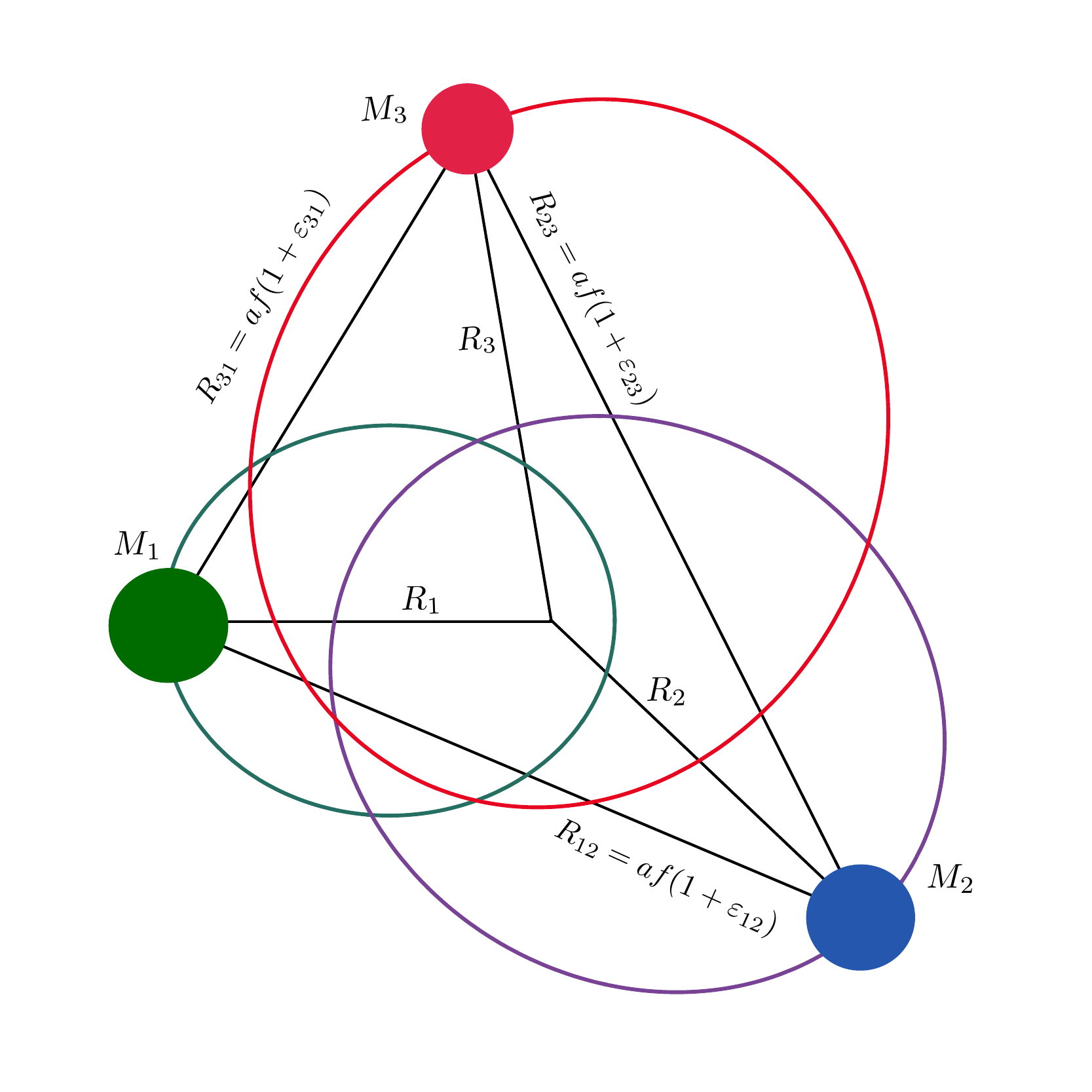}
\caption{
Schematic figure for the PPN triangular configuration of three masses. 
The inequilateral triangle is characterized by 
$\varepsilon_{AB}$. 
In the Newtonian limit, 
$\varepsilon_{AB}$ vanishes and 
$R_{AB}$ becomes  $a f_{\rm{N}}$.  
}
\label{figure-triangle}
\end{figure}

\section{PPN planar elliptic triangular solution}
\subsection{PPN planar elliptic orbit}
In order to obtain a PPN solution as a perturbation 
around the Newtonian 
equilateral elliptic solution, 
we assume a quasi-common dilation as 
$R_{AB} = a f (1 + \varepsilon_{AB})$ 
for three masses, 
where $\varepsilon_{AB}$ denotes a PPN distortion.  
The perfectly common dilation occurs at the Newton order, 
whereas the dilation is not common by $\varepsilon_{AB}$. 
See also Figure \ref{figure-triangle}.

In the same way as deriving Eq. (\ref{cross-Newton}), 
we take the cross product of $\bm{R}_1$ 
and Eq. (\ref{EOM-PPN})  for $M_1$ 
to obtain 
\begin{align}
&\ell_1^2 \frac{d}{dt} \bigg( f^2 \omega \bigg) 
(\bm{\lambda} \times \bm{\rho})
\notag\\
=& 
(\bm{\lambda} \times \bm{\rho})
\left\{
- \frac{\sqrt{3}}{2} \frac{M}{ f_{\rm{N}} a } \nu_2 \nu_3 
\right.
\notag\\
&
\left.
\times
\bigg[ 3  (\varepsilon_{12} - \varepsilon_{31}) 
+ \frac{M}{ 2 a }  ( \nu_3 - \nu_2 ) 
\bigg( \frac{1}{f_{\rm{N}}} - \frac{1}{\mathcal{A}_{\rm{N}}} \bigg) 
\right.
\notag\\
&~~~~
\left.
+ \frac{3}{8} a^2 \{\dot{f}_{\rm{N}} ( 1 + 3 \nu_1 ) 
+ \sqrt{3} f_{\rm{N}} \omega_{\rm{N}} (1 - \nu_1 - 2 \nu_2) \}
\right.
\notag\\
&~~~~~~~~~~
\left.
\times 
\{ \dot{f}_{\rm{N}} ( 1 - \nu_1 - 2 \nu_2 ) 
+ \sqrt{3} f_{\rm{N}} \omega_{\rm{N}} (1 - \nu_1) \} 
\right.
\notag\\
&~~~~~
\left.
- \frac{M}{4f_{\rm{N}}a} (\nu_2 - \nu_3) (8 \beta - 3)  \bigg] 
\right.
\notag\\
& - \frac{\sqrt{3}}{4} M a \nu_2 \bigg(
 \nu_3 \frac{\dot{f}_{\rm{N}}}{f_{\rm{N}}} 
+ \frac{\omega_{\rm{N}}}{\sqrt{3}} (\nu_1 - \nu_2 -1 ) \bigg)
\notag\\
&~~~~~
\left.
\times
 \bigg( ( 4 \gamma + 3 + \nu_2 - \nu_1 ) f_{\rm{N}} 
- \sqrt{3} \nu_3 f_{\rm{N}} \omega_{\rm{N}} \bigg) 
\right.
\notag\\
&
\left.
+ \frac{\sqrt{3}}{4} M a \nu_3 
\bigg( \nu_2 \frac{\dot{f}_{\rm{N}}}{f_{\rm{N}}} 
- \frac{\omega_{\rm{N}}}{\sqrt{3}} (  \nu_1 - \nu_3 - 1 ) \bigg)
\right.
\notag\\
&~~~~~
\left.
\times
\bigg( (4\gamma + 3 + \nu_3 -\nu_1) \dot{f}_{\rm{N}}
 + \sqrt{3} \nu_2 f_{\rm{N}} \omega_{\rm{N}} \bigg) 
 \right\} 
 \notag\\
 & + O(c^{-4}) ,
\label{ppneom}
\end{align}
where 
$\ell_1 \equiv a \sqrt{\nu_2^2 + \nu_2 \nu_3 + \nu_3^2}$ 
\cite{Danby, Roy, Yamada2012, Nakamura2023}, 
and we introduce an orthonormal basis $\bm{\lambda}$ and $\bm{\rho}$. 
Here,   
$\bm{\lambda} \equiv \bm{R}_1/R_1$,  
and $\bm{\rho}$ is the 90 degree rotation of $\bm{\lambda}$.
It is more convenient to use the orthonormal basis 
than $\bm{R}_1$ and $\bm{R}_2$, 
because the right-hand side of Eq. (\ref{EOM-PPN}) 
relies upon not only the positions but also the velocities. 
In elliptic motion, 
the velocity is not always orthogonal to the position vector, 
though it is in circular motion.

From the PPN total angular momentum, 
we find 
\begin{align}
& \frac{d}{dt} \bigg( f^2 \omega  \bigg) 
\notag\\
=& - \frac{M \dot{f}_{\rm{N}} \omega_{\rm{N}}}{4a} 
\frac{
13 \nu_1 \nu_2 \nu_3 - 8 \{
(\gamma + 1) \eta - \zeta 
\}}
{\eta} 
 + O(c^{-4}) , 
\label{dLdt}
\end{align}
where dot denotes the time derivative 
and 
we denote 
$\eta \equiv \nu_1 \nu_2 + \nu_2 \nu_3 + \nu_3 \nu_1$
and 
$\zeta \equiv \nu_1^2  \nu_2^2 + \nu_2^2  \nu_3^2 + \nu_3^2  \nu_1^2$. 
Eq. (\ref{dLdt}) is reduced to $d(f_N \omega_{\rm{N}}^2)/dt = 0$ in the Newtonian limit, 
which recovers the Newtonian case of the planar elliptic triangular solution. 
It follows that Eq. (\ref{dLdt}) can be derived also from the sum of 
Eq. (\ref{ppneom}) for $A = 1, 2, 3$. 

By substituting Eq. (\ref{dLdt}) into the left-hand side of Eq. (\ref{ppneom}), 
we obtain 
\begin{align}
&\varepsilon_{12} - \varepsilon_{31} 
\notag\\
=& \frac{M}{8 A} (\nu_3 - \nu_2) (3 \nu_1 + 1) 
\notag\\
&- \frac{M}{12 P} (\nu_3 - \nu_2) ( 9 \nu_1 + 8 \beta -2 ) 
( 1 + e_{\rm{N}} \cos\theta  ) 
 \notag\\
&+ \frac{M}{ 4 P} (\nu_3 - \nu_2) (3 \nu_1 - 1) 
( 1 + e_{\rm{N}} \cos\theta  )^2 
\notag\\
&
 - \frac{\sqrt{3}e_{\rm{N}} M}{72 \nu_2 \nu_3 P} 
 \sin\theta ( 1 + e_{\rm{N}} \cos\theta ) 
 \notag\\
& ~~~ \times 
\bigg[
34 \nu_1 \nu_2 \nu_3 + 16 \nu_1 (\nu_2^2 + \nu_3^2) 
\notag\\
&~~~~~
+ 9 \nu_2 \nu_3 \{ 1 - 3 \nu_1^2 + (\nu_2 - \nu_3)^2 \} \nonumber\\
&~~~~~ 
- \frac{4 (\nu_2^2 + \nu_2 \nu_3 + \nu_3^2)(13 \nu_1 \nu_2 \nu_3 + 8 \zeta)}{\eta} 
\bigg] + O(c^{-4}), 
\label{epsilon12-31}
\end{align}
where Eq. (\ref{orbit-Newton}) is used for $f_N$. 
By cyclic arguments, 
$\varepsilon_{23} - \varepsilon_{12}$
and 
$\varepsilon_{31} - \varepsilon_{23}$ 
are obtained.

Following Reference 
\cite{Yamada2012}, 
the gauge fixing is chosen as 
$\varepsilon_{12} + \varepsilon_{23} + \varepsilon_{31} = 0$, 
for which the PN triangular area remains the same as the Newtonian one. 
From this gauge fixing and Eq. (\ref{epsilon12-31}), 
we obtain 
\begin{align}
&\varepsilon_{12} 
\notag\\
=& 
\frac{M}{24 A} [3 \{ \nu_1 (\nu_3 - 2 \nu_2) +\nu_3 (1 + \nu_2)\} - 1] 
\notag\\
&- \frac{M}{36 P} 
[2 (4 \beta - 1) ( 3 \nu_3 - 1) 
\notag\\
& ~~~~
+ 9 \{ \nu_1 (\nu_3 - \nu_2) + \nu_2 (\nu_3 - \nu_1) \}] 
( 1 + e_{\rm{N}} \cos \theta ) 
\notag\\
&+ \frac{M}{ 12 P}  [1 -3 (\nu_3^2 + 2 \nu_1 \nu_2)] 
( 1 + e_{\rm{N}} \cos \theta  )^2 
\notag\\
&- \frac{e_{\rm{N}} M \sqrt{3} }{108 P} (\nu_1 - \nu_2) 
\sin\theta  ( 1 + e_{\rm{N}} \cos \theta )
\notag\\
& ~~~ \times 
\bigg[
\frac{8 \nu_3 (1 - \nu_3)}{\nu_1 \nu_2} + 27 \nu_3 - 1 
\notag\\
& ~~~~~~~~
+ \frac{2 (\nu_1 \nu_2 - \nu_3^2) (13 \nu_1 \nu_2 \nu_3 + 8 \zeta)}{\nu_1 \nu_2 \nu_3 \eta} 
\bigg] 
 + O(c^{-4}).
\label{varepsilon12}
\end{align}

It is worthwhile to mention that 
$\gamma$ makes no contribution to Eq. (\ref{varepsilon12}), 
while $\beta$ is included in it. 
This means that 
the PPN nonliearity parameter $\beta$ affects the asymmetric shape of the PN triangle,  
whereas $\gamma$ does not affect the asymmetry.

We can obtain also 
$\varepsilon_{23}$ and $\varepsilon_{31}$ cyclically. 
We thus obtain the PPN triangular quasi-elliptic solution. 
Note that this solution does not follow a perfectly elliptic motion 
owing to the periastron shift as shown below, 
whereas
the Newtonian counterpart is elliptic. 
For instance, the periastron position in the PPN orbit 
moves significantly in a long time scale. 
Namely,  the obtained solution represents an osculating orbit 
\cite{Danby,Roy}.

\begin{figure}
\includegraphics[width=8.0cm]{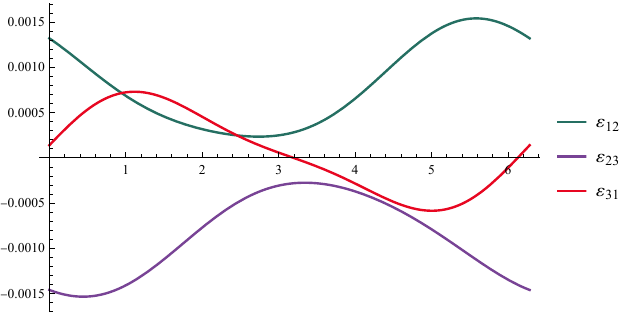}
\caption{
$\varepsilon_{12}, \varepsilon_{23}, \varepsilon_{31}$ 
for $\nu_1 = 1/2$, $\nu_2 = 1/3$, $\nu_3 = 1/6$ 
and $M/a = 0.01$ 
in elliptic motion with $e_{\rm N} = 0.5$.
The horizontal axis denotes $\theta$ 
from a periastron ($\theta = 0$) 
to the next periastron ($\theta = 2\pi$). 
}
\label{fig-Psi}
\end{figure}

\subsection{Periastron shift}
After direct calculations, 
the PPN expression of the total energy 
\cite{Baker1978,Baker1979} 
for the PPN planar quasi-elliptic triangular solution 
can be rewritten as 
\begin{align}
\bigg( \frac{d u}{d \theta}  \bigg)^2+ G(u) \frac{d u}{d \theta} = F(u) , 
\label{orbiteq}
\end{align}
where $u \equiv 1/f$.   
$F(u)$ and $G(u)$ are functions of $u$, 
which are too long to write down in this paper. 

The periastron shift is 
\begin{align}
\theta_{\rm{PPN}}  
= 
\int_{u_{\rm{min}}}^{u_{\rm{max}}}
du 
\frac{1}{\left(\cfrac{du}{d\theta}\right)} 
- \pi , 
\label{shift}
\end{align}
where $u_{\rm{max}}$ and $u_{\rm{min}}$ 
correspond to the apoapsis 
and periapsis, respectively.

In the same way as the post-Newtonian calculations of the periastron shift 
\cite{Will,Poisson}, 
by using Eq(\ref{orbiteq}) for Eq. (\ref{shift}),  
we obtain 
the periastron shift at the PPN order as 
\begin{align}
&\theta_{\rm{PPN}}
 \notag\\
 =&
\frac{\pi M}{36 P \eta} 
\bigg[
18 \nu_1 \nu_2 \nu_3 (9 - 2 \beta) 
+ \eta (65 - 44 \beta + 72 \gamma)
+ 36 \zeta  
\bigg] 
\notag\\
 & + O(c^{-4}) .
 \label{thetaPPN}
\end{align}
The periastron shift per orbital period is $2\theta_{\rm{PPN}}$. 
In GR ($\beta = \gamma = 1$), 
Eq. (\ref{thetaPPN}) becomes 
\begin{align}
\theta_{\rm{PPN}}
=&
\frac{\pi M}{36 P \eta} 
(126 \nu_1 \nu_2 \nu_3 
+ 93 \eta 
+ 36 \zeta) 
 + O(c^{-4}).
 \label{thetaGR}
\end{align}

In the test particle limit of a third mass ($\nu_3 \to 0$), 
Eq. (\ref{thetaPPN}) disagrees with 
that of a binary system, 
because the restricted three-body dynamics  
does not equal to the binary dynamics 
\cite{Danby,Roy,Yamada2012}. 
See e.g. 
Eq. (66) in Reference \cite{Will}
and 
Eq. (13.51) in \cite{Poisson} 
for the PPN periastron shift formula of a binary case.


\section{Conclusion}
We found a PPN triangular solution 
to the planar elliptic three-body problem 
in a class of fully conservative theories. 
The distortion function $\varepsilon_{AB}$ 
of a triangular solution 
depends on $\beta$ but not on $\gamma$. 
It follows that, 
in the circular limit,  
the present solution recovers the PPN triangular circular solution 
in Reference \cite{Nakamura2023}. 
In the limit of $e_{\rm{N}} \to 0$, 
Eq. (\ref{varepsilon12}) agrees with 
Eq. (41) in \cite{Nakamura2023}. 

The periastron shift of the PPN triangular solution 
was also obtained. 
Because of the three-body interactions, 
the periastron shift in the PPN triangular solution 
is different from that of a binary system.

There are potential observational tests for the above models. 
One is the monitoring of an artificial satellite 
at (or around) $L_4$ (or $L_5$) 
of the Sun-Jupiter system (or Sun-Earth system), 
if such a satellite is launched. 
It could allow to test the relativistic three-body gravity through the measurement of  
$\beta$ and $\gamma$, though it is technically difficult. 

The other is to find a hypothetical object of a relativistic triangular system 
composed from e.g. two black holes and a neutron star. 
If the two black holes are much heavier than the neutron star, 
the triple system is likely to be stable, 
though its formation process is unclear.

It is left for future to study the stability of the present solution.

\section{Acknowledgments}
We thank Yuuiti Sendouda and Marcus Werner 
for encouraging comments. 
This work was supported 
in part by Japan Science and Technology Agency (JST) SPRING, 
Grant Number, JPMJSP2152 (Y.N.), 
and 
in part by Japan Society for the Promotion of Science (JSPS) 
Grant-in-Aid for Scientific Research, 
No. 20K03963 (H.A.).


\begin{thebibliography}{99}
\bibitem{Goldstein}
H. Goldstein, {\it Classical Mechanics} 
(Addison-Wesley, MA, 1980). 
\bibitem{Danby}
J. M. A. Danby, {\it Fundamentals of Celestial Mechanics} 
(William-Bell, VA, 1988). 
\bibitem{Roy}
A. E. Roy, 
{\it Orbital Motion} 
(Adam Hilger, Bristol, 1982). 
\bibitem{Marchal}
C. Marchal, {\it The Three-Body Problem} 
(Elsevier, Amsterdam, 1990). 
\bibitem{Asada}
H. Asada, Phys. Rev. D {\bf 80} 064021 (2009). 
\bibitem{Torigoe}
Y.~Torigoe, K.~Hattori and H.~Asada, 
Phys.\ Rev.\ Lett.\  {\bf 102}, 251101 (2009).
\bibitem{Seto}
N. Seto, T. Muto, Phys. Rev. D {\bf 81} 103004 (2010).
\bibitem{Schnittman}
J. D. Schnittman, Astrophys. J. {\bf 724} 39 (2010). 
\bibitem{Connors}
M. Connors, P. Wiegert and C. Veillet, 
Nature {\bf 475}, 481 (2011).
\bibitem{Nordtvedt}
K. Nordtvedt,  
Phys. Rev. {\bf 169} 1014 (1968).
\bibitem{Krefetz}
E. Krefetz, Astron. J {\bf 72}, 471 (1967).
\bibitem{Maindl}
T. I. Maindl, 
{\it Completing the Inventory of the Solar System, Astronomical Society of
the Pacific Conference Proceedings}, edited by T.W. Rettig and
J.M. Hahn, (Astronomical Society of the Pacific, San Francisco, 1996), 107, 147.
\bibitem{Yamada2010}
K. Yamada, H. Asada, 
Phys. Rev. D {\bf 82}, 104019 (2010).
\bibitem{Yamada2011}
K. Yamada, H. Asada, 
Phys. Rev. D {\bf 83}, 024040 (2011).
\bibitem{Ichita2011}
T. Ichita, K. Yamada, H. Asada, 
Phys. Rev. D {\bf 83}, 084026 (2011).
\bibitem{Yamada2012}
K. Yamada and H. Asada, 
Phys. Rev. D {\bf 86}, 124029 (2012).
\bibitem{Yamada2015}
K. Yamada, T. Tsuchiya and H. Asada
Phys. Rev. D {\bf 91}, 124016 (2015).
\bibitem{Yamada2016}
K. Yamada and H. Asada, 
Phys. Rev. D {\bf 93}, 084027 (2016).
\bibitem{Zhou}
T. Y. Zhou, W. G. Cao, and Y. Xie, 
Phys. Rev. D {\bf 93}, 064065 (2016). 
\bibitem{Battista2015a}
E. Battista, S. Dell'Agnello, G. Esposito, and J. Simo, 
Phys. Rev. D {\bf 91}, 084041 (2015); 
Erratum, Phys. Rev. D {\bf 93}, 049902(E) (2016).
\bibitem{Battista2015b}
E. Battista, S. Dell’Agnello, G. Esposito, L. Di Fiore, J. Simo, and A. Grado, 
Phys. Rev. D {\bf 92}, 064045 (2015);
Erratum, Phys. Rev. D {\bf 93}, 109904(E) (2016).
\bibitem{Ransom}
S. M. Ransom, I. H. Stairs, A. M. Archibald, J. W. T. Hessels, D. L. Kaplan 
and et al.
Nature, {\bf 505}, 520 (2014).
\bibitem{Archibald}
Anne M. Archibald, Nina V. Gusinskaia, Jason W. T. Hessels, 
Adam T. Deller, David L. Kaplan and et al.
Nature, {\bf 559}, 73 (2018).
\bibitem{Will2018} 
C.~M.Will, Nature, {\bf 559}, 40 (2018). 
\bibitem{Voisin}
G. Voisin, I. Cognard, P.~C.~C. Freire,  et al., 
Astron. Astrophys. {\bf 638}, A24 (2020).
\bibitem{Will}
C. M. Will, Living Rev. Relativity, {\bf 17}, 4 (2014). 
\bibitem{Poisson}
E. Poisson, and C. M. Will, 
{\it Gravity}, (Cambridge Univ. Press, UK. 2014). 
\bibitem{Klioner}
S. A. Klioner and M. H. Soffel, 
Phys. Rev. D {\bf 62}, 024019 (2000).
\bibitem{Baker1978}
B. M. Barker and R. F. O'Connell, 
Phys. Lett. A {\bf 68}, 289 (1978). 
\bibitem{Baker1979}
B. M. Barker and R. F. O'Connell, 
J. Math. Phys. {\bf 20}, 1427 (1979).
\bibitem{MTW}
C. W. Misner, K. S. Thorne, and J. A. Wheeler, 
{\it Gravitation} 
(Freeman, New York, 1973).
\bibitem{LL}
L. D. Landau and E. M. Lifshitz, 
{\it The Classical Theory of Fields} 
(Pergamon, New York, 1962).
\bibitem{Nakamura2023}
Yuya Nakamura and Hideki Asada, 
Phys. Rev. D 107, 044005 (2023).
\end{thebibliography}
\end{document}